\documentclass{aastex}
\usepackage{spr-astr-addons}
\usepackage{url}

\RequirePackage{color}

\newcommand{\emaila}{hcakmak@istanbul.edu.tr}

\usepackage{graphicx}
\usepackage{stfloats}
\usepackage[outline]{contour}

\newcommand{\quotes}[1]{``#1''}

\newcommand{\exas}{Exp. Astron.} 
\newcommand{\jastp}{J. Atmos. Solar-Terr. Phys.}

\begin{document}

\title{A semi-automated method to reveal the evolution of each sunspot group in a solar cycle}
\slugcomment{Not to appear in Nonlearned J., 45.}
%% Running heads
\shorttitle{A semi-automated method to reveal the evolution of each sunspot group}
\shortauthors{H. {\c C}akmak}

\author{H. {\c C}akmak\altaffilmark{1}} 

\email{\emaila}

\altaffiltext{1}{Istanbul University, Faculty of Science, Department of Astronomy
	and Space Sciences, 34116, University, Istanbul, Turkey}

\begin{abstract}
Sunspots are the most important indicator of the magnetic activity on the solar surface during a cycle. Every sunspot group is formed and shaped by the magnetic field of the Sun. Hence, the magnetic field intensity shows itself as the size of a sunspot group area on the surface. This shows that getting the development or evolution of sunspot groups over time means getting the change of magnetic field intensity during same interval. Here, to reveal the evolution of sunspot groups in a cycle, a method called {\it Solar Cycle Analyzer Tool} (SCAT) is presented. This method was developed as a part of {\it Computer-Aided Measurements for Sunspots} (CAMS) because the same subroutines and subprograms were used for calculations \citep{Cakmak2014}. The developed software tracks sunspot groups every day and gives them the same group number. The confirmation is made by the user to prevent counting re-formations as a continuation of an old group in the same active region. With this method, the evolution of every sunspot group can be listed for each cycle year besides other cycle features like the daily and monthly sunspot relative numbers and distribution frequency of the sunspot group types. Since 2015, SCAT is being used to get data for the annual reports of Istanbul University Observatory.
\end{abstract}


\section{Introduction} \label{S-intro}

\begin{figure}[!b]\label{F-figures-01}
	\centerline{\includegraphics[width=1.0\columnwidth]{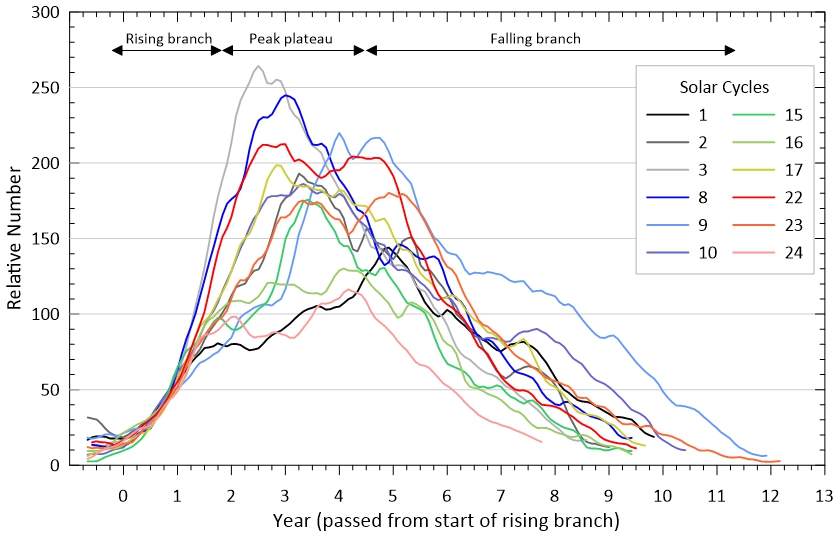}}
	\caption{Comparison of several solar cycles. Rising branches of every solar cycle are arbitrarily shifted so they all coincide approximately at the same point to show the differences of maximums and widths regarding each other properly. The lines with arrows on both ends on top show the sections of a solar cycle profile. Sunspot data are taken from the {\it World Data Center SILSO, Royal Observatory of Belgium, Brussels}.}
\end{figure}

Sunspot cycles are continuing to keep their mystery. We are now in the 24th sunspot cycle, which is ending. Every cycle shows a different development profile in a magnetic activity expressed by the sunspot relative number. Although a general Gaussian profile is observed in these development profiles, the maximum and width of the profiles change non-periodically from cycle to cycle (Fig. 1). This non-periodicity makes it difficult to find clues about underlying mechanisms that cause sunspot formation on the surface of the Sun. When all cycle profiles are taken into account one can see that every cycle profile has three different sections that can be named as {\it rising branch}, {\it peak plateau} and, {\it falling branch}. Each part of the profile has a different behavior regarding its progressive shape and duration. These differences mainly arise from the rate of increase or decrease in the number of sunspot groups formed in each section during the cycle. One can think the prominent feature about cycle progress would be the monthly sunspot formation rate. In advanced studies about cycles, one may switch this feature to weekly or daily sunspot formation rate depending on which branch of the cycle is under consideration to reveal section properties. 

\begin{figure*}[!t]\label{F-figures-02}
	\centerline{\includegraphics[width=1.0\textwidth]{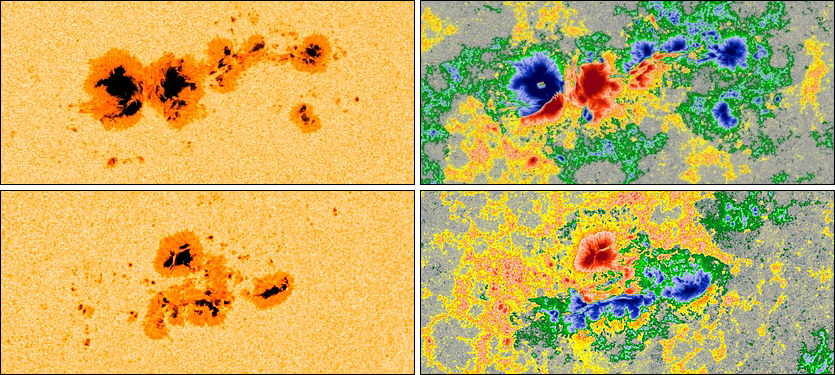}}
	\vspace{-0.445\textwidth} 
	\hspace{0.0\textwidth} \color{black} (a) \\
	
	\vspace{0.133\textwidth} \contourlength{.3pt}
	\hspace{0.0\textwidth}  \textcolor{white}{\contour{black}{HMIIF - Visual}} 
	\hspace{0.36\textwidth} \textcolor{white}{\contour{black}{HMIBC - Magnetogram}}\\
	%%\color{black} HMIBC - Magnetogram
	
	\vspace{-0.01\textwidth} 
	\hspace{0.0\textwidth} \color{black} (b) 
	
	\vspace{0.18\textwidth}
	\caption{Two samples of entangled (or intertwined multiple) sunspot groups observed on \textbf{a)} 3 February 2014 and \textbf{b)} 19 December 2014 during the maximum phase of 24th Solar Cyle. Chaotic structure of the sunspot groups is more clear in magnetogram images on the right panel. Courtesy of NASA/SDO and the AIA, EVE, and HMI science teams..}
\end{figure*}

The observed types of sunspot groups show a great deal of variety from the beginning to the end of the cycle. Relatively small sunspot groups emerge at the beginning and as time goes by more developed and complex groups appear during the cycle maximum. When the cycle period reaches its maximum, more sophisticated and chaotic sunspot groups appear frequently as observed in most of the cycles \citep{BhatLiv2005,Lang2009}. These development patterns of sunspot groups were classified by \citet{Waldmeier1947} using a method he named as {\it Zurich Sunspot Classification}. With the help of increasing sunspot observations, \citet{McIntosh1990} further developed this classification by using the patrol observations of Space Environment Services Center\footnote{The service portion of the Space Environment Laboratory of the U.S. National Oceanic and Atmospheric Administration.} made between the years 1960 and 1976. This classification is nowadays referred to as {\it McIntosh Sunspot Group classification}. Although such classifications provide great contributions to reveal the formation and development of the sunspot groups, there are still some difficulties in classification of some observed sunspot groups, especially for entangled (or intertwined multiple) sunspot groups (Fig. 2). Encountering such examples frequently shows that new arrangements and definitions in sunspot classification should be done properly to classify these ambiguous groups. Therefore, it is necessary to follow up as many entangled sunspot groups as possible. Information about new arrangements for these groups can be obtained by analyzing their evolution. Disclosing the evolution of a sunspot group has a special importance for this perspective. Questions such as how and at what stage the sunspot groups become complicated or entangled will be answered by examining these evolution stages. From this point of view, SCAT provides great convenience for data processing and it reveals not only the group evolution easily but also provides some statistical information about the cycle in question.

SCAT is mainly constructed on the premise that each group in the cycle must have a unique group number. This provides an opportunity to distinguish every sunspot group from each other easily in programmatic approach. Also, other information such as average latitude and longitude, lifetime and first and last observation dates of the group are automatically collected during data processing. SCAT gives some informational data about the processed cycle such as daily sunspot relative number in a list for the whole year and monthly average relative number and the list of evolution of each sunspot group for the cycle year in question. Nowadays, there are no programs or methods in the literature which gives this kind of information. Methods mostly give heliographic coordinates of the sunspot groups with solar parameters on the observation day. The most commonly known applications are the Helio programs developed by Peter \cite{Meadows2002}. The other one HSUNSPOTS is developed by \cite{Cristo2011} to analyze ancient solar drawings and DigiSun is used in the SIDC of the Royal Observatory of Belgium \citep{Clette2011}.

The methodology of tracking a sunspot group every day is given in Section 2 along with the equations. The heliographic coordinates ({\it B}, {\it L}) of the sunspot group is used to check their displacement between the checked days during tracking process. The working scheme of SCAT program is introduced and some screenshots are given to show its general perspective in Section 3. The results of a working example are given in Section 4 along with relevant data. In Conclusion, possible developments of the program are discussed with a future perspective.

\begin{figure}[!b]\label{F-figures-03}
	\centerline{\includegraphics[width=0.85\columnwidth]{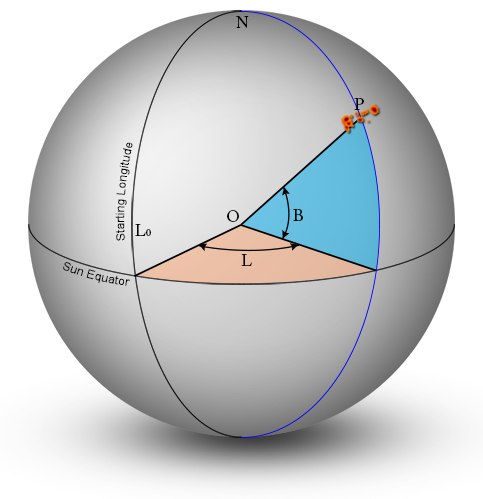}}
	\caption{Schematic representation of the heliographic coordinates of a sunspot group. {\it O} is the center of the Sun, {\it N} is the North pole, {\it $L_0$} is the starting (or zero) longitude and, {\it P} is the position of the sunspot group on the solar surface.}
\end{figure}

\section{Tracking A Sunspot Group} \label{S-track}
Every sunspot group on the surface of the Sun has heliographic coordinates represented with latitude {\it B} and longitude {\it L}. A sample group is shown in Fig. 3 \textendash \, general information and detailed explanations about this coordinate system are given in the article of {\c C}akmak published in 2014. Since the heliographic coordinate system is specially designed for the Sun’s surface, these coordinates show the location of a group on the surface precisely except the groups with great changes in their length. Basically, the heliographic coordinate of a group does not change much during its development. So, the changes in coordinates remain within a few degrees, and they are tiny compared to the group’s length. This small displacement facilitates the following up of a group on consecutive days. The heliographic coordinate differences between the considered group and other groups in previous days will be at the minimum value at the same time in latitude and longitude. Hence, it is necessary to create a checklist of the groups observed in previous days to track a group properly. Also, this list must be wide enough to compare all groups of a day on the visible surface of the Sun. 

When a disk passage of a group is taken into account, one can see that fastest transition lasts approximately 13 days in the equatorial region for the visible surface of the Sun. But due to the solar differential rotation, this transition period varies depending on the group’s latitude and increases as the group approaches the solar poles \citep{Howard1984,Balthasar1986}. In the first approach, total day number for the list was accepted as 17 days by considering the last observation day of a sunspot which comes from the back side of the Sun and becomes visible on the eastern part of the solar disk. Because of some experiments made during program development, this number is changed to 31 days by considering all conditions that the sunspot observations can not be made because of cloudy and rainy days, observatory maintenance and telescope malfunctions. In these cases, sunspot observations will be interrupted and the number of unobserved days will increase. As an inevitable consequence of this, it is necessary to go back further in a past to skip these empty days and find the last observation day for the proper match. So, when this 31-day group checklist is obtained, any group outside can be found easily by scanning this list starting from the closest day to the farthest day.

When a group is compared with another, both groups are supposed to be on the same solar disk. Now, let us have two groups that are the same group in two different days and let latitude and longitude coordinates of these groups be $B_S$, $L_S$, $B_C$ and $L_C$, respectively. Also, let latitudinal and longitudinal lengths of them be $H_S$, $W_S$, $H_C$ and $W_C$, respectively. Since both groups are the same group, their positions on the solar disk will be very close and sometimes may be superimposed completely. A sample of this situation is shown exaggeratedly with real sunspot images\footnote{These images are taken from the SDO archive on the date of 15 and 16 June 2012.} in Fig. 4 in order to  show clearly the used parameters for calculation. Here, each group is represented with a rectangle showing width and height of the group’s area and center of the rectangle (blue points) shows the latitude and longitude of the group. Let these center points be $P_S$ and $P_C$, respectively. 

\begin{figure}[t!]\label{F-figures-04}
	\centerline{\includegraphics[width=1.0\columnwidth]{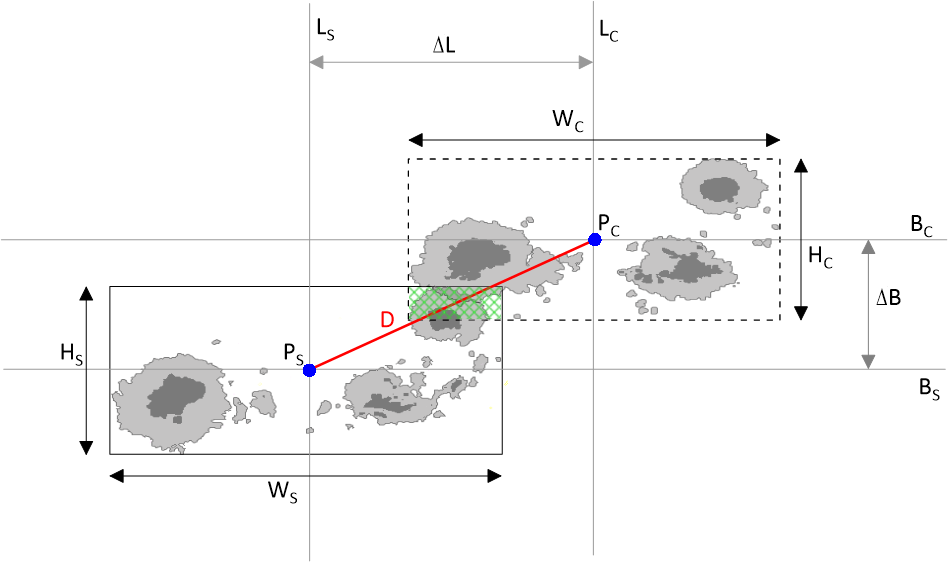}}
	\caption{The comparison parameters of two sunspot groups to calculate the group's proximity.}
\end{figure}

Two parameters should be taken into account for comparison under the condition that both groups have to be in the same hemisphere. One is the distance of group centers, which is shown with a red line in Fig. 4, and other is the amount of overlap between two group areas, which is shown with green grid area in Fig. 4. The distance $D$ between group centers in degrees is calculated by \citep{Green1986}
\begin{flalign}  \label{Eq-distance}   
\Delta B &= \vert B_S - B_C \vert \,, \nonumber  \\[0pt]
\Delta L &= \vert L_S - L_C \vert \,, \nonumber  \\[0pt]
D &= \arccos \,(\sin B_S \sin B_C + \cos B_S \cos B_C \cos \Delta L) 
\end{flalign}
where $\Delta B$ and $\Delta L$ are the absolute differences between the latitudes and longitudes of both groups, respectively. As seen in Fig. 4, the acceptable maximum distance value ($D_0$) can be calculated by using the half widths and half heights of the group areas. $D_0$ is given by
\begin{flalign}  \label{Eq-dist-lat} 
D_{B0} &= A \times \bigg[ \frac{H_S + H_C}{2} \bigg] \,, D_{L0} = B \times \bigg[ \frac{W_S + W_C}{2} \bigg] \,,  \\[4pt]
D_0 &= \sqrt{(D_{B0})^2 + (D_{L0})^2} 
\end{flalign}
where $D_{B0}$ is the sum of half heights and $D_{L0}$ the sum of half widths of both group areas. $A$ and $B$ are the coefficients used to change the contribution rate of width or height to the maximum distance. When Equation 2 is analyzed carefully, one see that these coefficients are directly giving the amount of overlap in latitude and longitude separately, because the value of 1 for a coefficient shows that both groups are next to each other and they are not overlapped. Only the numbers less than zero show the overlapped situation for the groups. With various trials made using real data, $A$ and $B$ coefficients are taken as 1.12 and 0.84, respectively. Hence, the following terms must be valid at the same time to find group’s pair exactly in the past groups list:
\begin{flalign}  \label{Eq-condit}   
D  \leqslant D_0 \,,  \quad\quad \Delta B  \leqslant D_{B0} \,, \quad\quad  \Delta L  \leqslant D_{L0} \,.
\end{flalign}
When these terms are valid for a group, the group number for a new group is inherited from the past so that group evolution has been followed correctly. If no group pair was found in the past groups list, a new group number is assigned to a new group by increasing the last group number by one unit. Also, aforementioned terms provide an opportunity to find the sunspots that make their second, third or more turns on the solar surface due to being on the same active region.

\begin{figure}[b!]\label{F-figures-05}
	\centerline{\includegraphics[width=0.99\columnwidth]{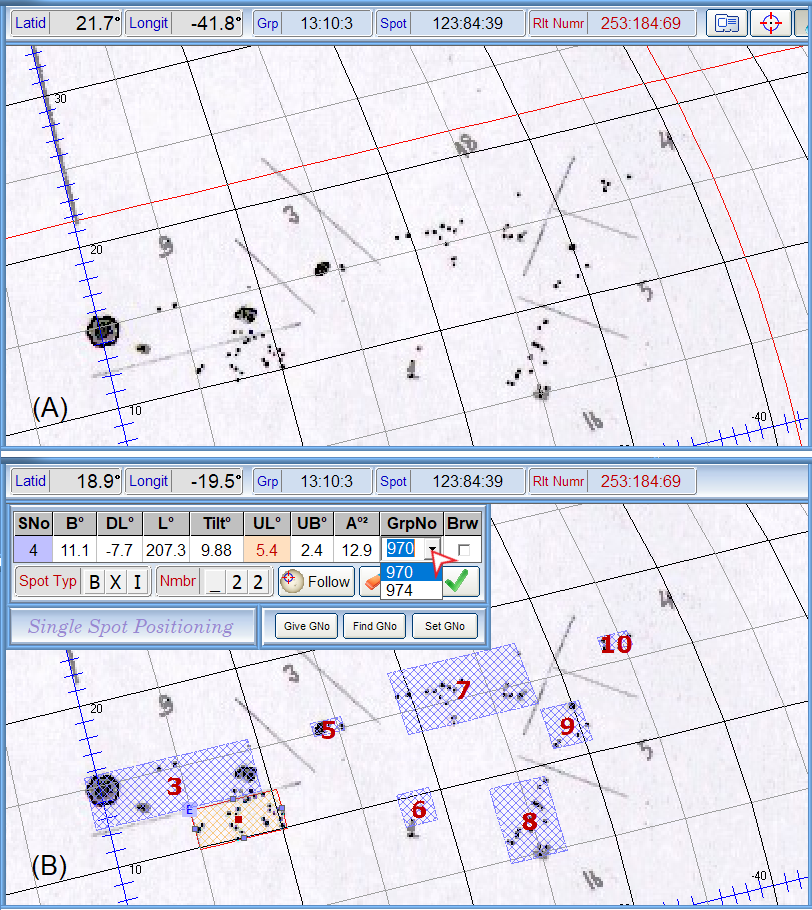}}
	\caption{An example of sunspot groups that are close together observed on June 5, 2012. \textbf{A)} Appearance of the groups before separation. \textbf{B)} Group numbering process after the groups are manually separated by using CAMS.}
\end{figure}

Along with not being observed often, there are sunspot groups which are close to each other and distinguishing them from each other will not easily be possible in a programmatic approach. Therefore, the list of possible group numbers is created as a solution for such groups by scanning whole past groups list. A sample is shown in Fig. 5-B on a daily solar drawing taken on June 5, 2012, at Istanbul University Observatory. In such observations, all sunspot groups are separated manually in the CAMS program. But, as known, the group separation depends entirely on the people's preference. Therefore, a brief description about this subject will be given in Discussion section. In such cases where groups are close, the user must select a proper group number by analyzing the group developments history, i.e. 31-day group control list. This is the reason why this method is called as semi-automated.

\section{Working Scheme of the Method} \label{S-scheme}
As mentioned in the introduction section, the developed method is a part of the CAMS \citep{Cakmak2014}. Basically, the method is comprised of two sections: the first is the assignment of group numbers to the whole sunspot groups in the considered year and the second is the yearly analyzing. Before starting the first stage, the initial group number must be adjusted according to whether it is inherited from last group number in the program database or it is given a new value. Importantly, it must be specified that giving group number process has started by activating its check-box (Fig. 6). A previously processed sunspot group without a group number is reloaded into CAMS, a 31-day group checklist will be created automatically from the program database, and a group number list containing the appropriate group numbers is shown in the group information window \textemdash small window on top of the disk window in Fig. 7.

\begin{figure*}[b!]\label{F-figures-06}
	\centerline{\includegraphics[width=0.85\textwidth,clip=]{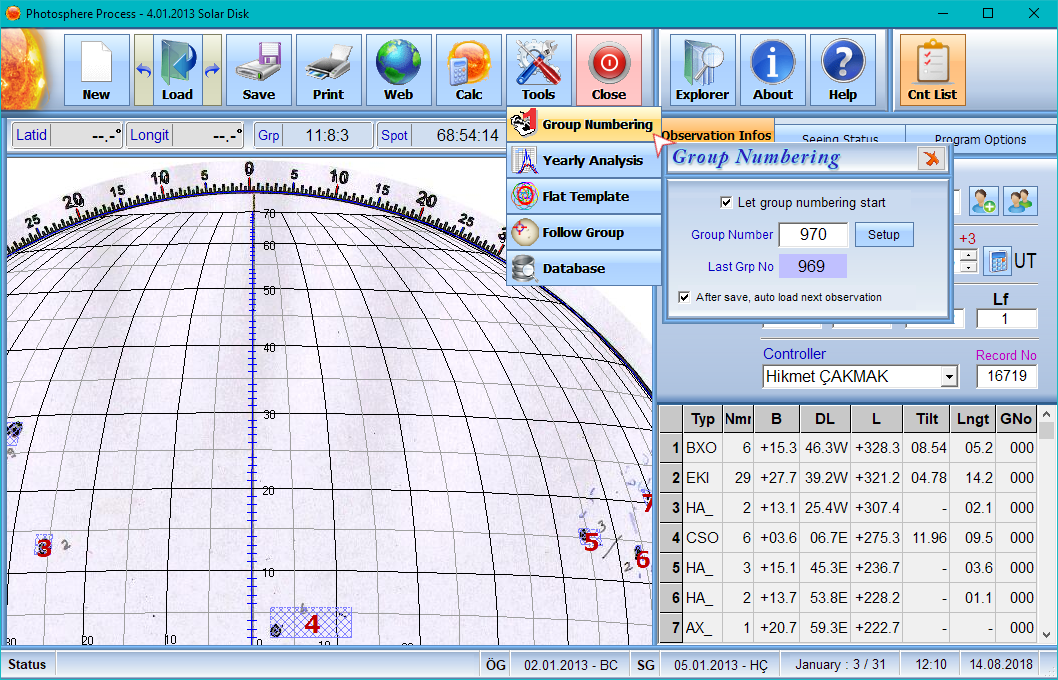}}
	\caption{The adjustment of the group numbering initialization.}
\end{figure*}

If the selected sunspot group is a new group of the present day, which has appeared on this day, i.e., it has no history, the new group number will be given by increasing the used last group number by one, and the heritage check-box will be un-ticked in this case showing the new group status. After confirmation is done, the observation image of the next day will automatically be loaded and the first sunspot group of the day will also be selected automatically for processing.

If the selected sunspot group has a history, i.e., if it has appeared on the solar disk earlier, the group number of the previous sunspot group will be selected under the aforementioned criteria (Equ. 4). Here, the heritage check-box which represents that the group number is inherited or borrowed from previous group will be automatically activated in the group information window. When the 31-day group control list is opened by clicking its button at this stage, all previous groups belong the selected group will be highlighted with a different color to track them easily in the list, and the observation image of the considered day will also be loaded. This situation is shown in Fig. 8 with an example. As seen this figure, all group numbers are shown over related sunspot groups on the observation image of the previous day. This makes it easier to not only to check the group number, but also to check the group position on the solar disk visually.

\begin{figure*}[!t]\label{F-figures-07}
	\vspace*{5mm}
	\centerline{\includegraphics[width=0.85\textwidth]{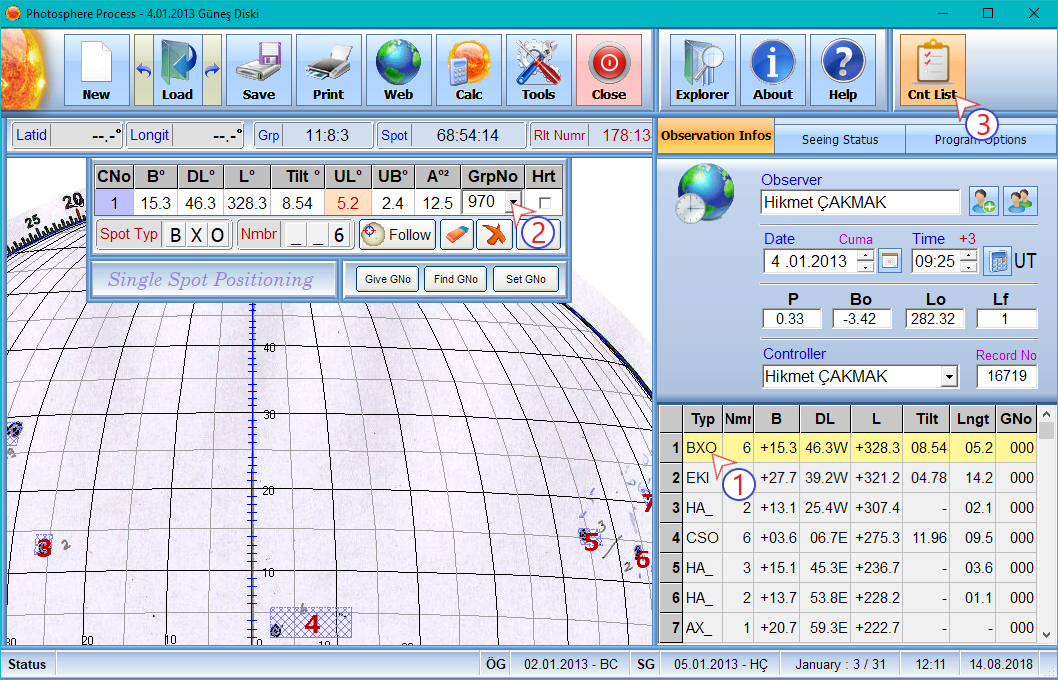}}
	\caption{Group numbering stages. \textbf{1)} Selecting a sunspot group that has no group number, \textbf{2)} choosing a proper group number, \textbf{3)} checking the previous group developments in case of hesitation.}
\end{figure*}

\begin{figure*}[h!]\label{F-figures-08}
	\vspace*{5mm}
	\centerline{\includegraphics[width=0.85\textwidth]{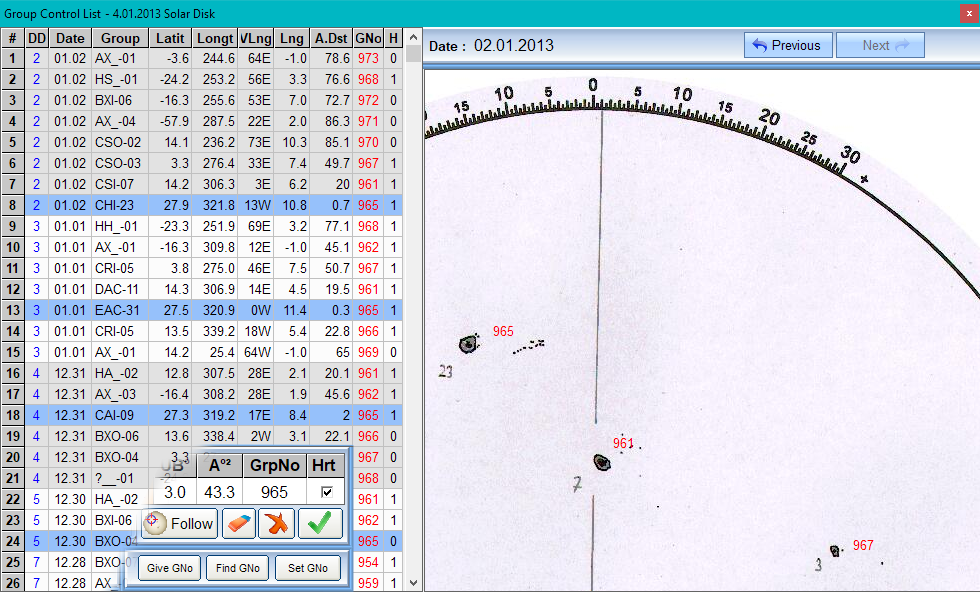}}
	\caption{31-day sunspot group control list with an example of highlighted group. This situation is only valid for the sunspot group which its group number is inherited from previous sunspot group as shown on left-bottom.}
\end{figure*}

When assigning the group numbers to sunspot groups, some special or extraordinary situations may be encountered, which cause confusion in group numbers such as being not sequential or using same number more than one. These cases can be solved by making an appropriate choice of group numbering buttons named as {\it Give GNo}, {\it Find GNo} and {\it Set GNo} in group information window (Fig. 7). The used last group number can be given by clicking \quotes{Give GNo} button, and the last group number taken from program database can be given by \quotes{Find GNo}. But if any group number is omitted or not used, \quotes{Set GNo} button must be used. A sample for this case is shown in Fig. 9. First, the sunspot group before the unused group number should be selected even though it is on another day. Then, all group numbers after this group should be increased or decreased by one according to the selection wanted. So, replacement operations are performed after the desired button is pressed.
\begin{figure}[!t]\label{F-figures-09}
	\centerline{\includegraphics[width=1.0\columnwidth]{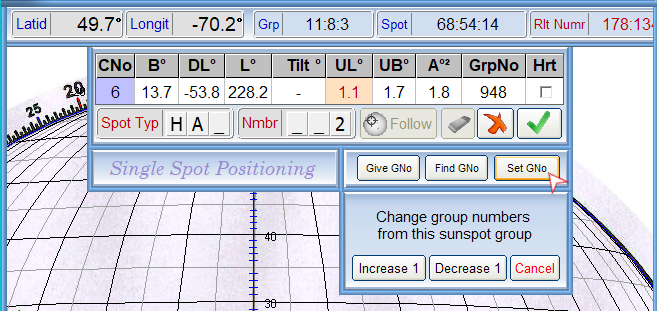}}
	\caption{Changing the whole group numbers starting from a group that group number sequence is broken.}
\end{figure}

\begin{figure*}[!b]\label{F-figures-10}
	\centerline{\includegraphics[width=0.80\textwidth]{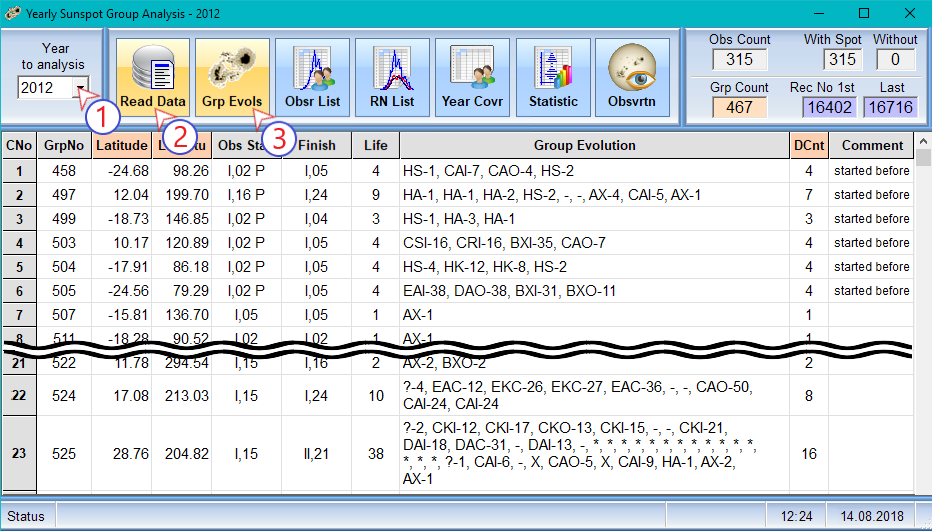}}
	\caption{Steps of a sample process in the Yearly Analysis section. \textbf{1)} Selecting a year to be analyzed, \textbf{2)} reading all data of the selected year from program database, \textbf{3)} extracting the evolution of whole sunspot groups separately, as an example.}
\end{figure*}

When the group numbers of all sunspot groups in considered year are given with no conflict, the second stage of the developed method is started by selecting the \quotes{Yearly Analysis} from Tools menu of the program. This section is generally designed to meet the needs of Istanbul University Observatory, especially for the observatory annual report. The results of some specific issues about the cycle year in question are given in this part of program, and all results in each subsection are obtained with just one click. The general outlook of this part is shown in Fig. 10 with the result of a sample process. As seen from this figure, there are six different analysis sections on the program menu. These are named as {\it Group Evolution} (Sec-A), {\it Observer List} (Sec-B), {\it Relative Number List} (Sec-C), {\it Year Cover} (Sec-D), {\it Statistic} (Sec-E) and {\it Observations} (Sec-F), respectively.

In Sec-A, the evolution of whole sunspot groups is given as a list according to the order of appearance on the visible solar disk. A demonstration is given in Fig. 10. The obtained properties of the sunspot groups in this list are as follows; average latitude and longitude, first and last observation date as month-day, visible life\footnote{This time interval is calculated from the first and last observation day of the sunspot group. If the end of the lifetime of a sunspot group is taken place on the backside of solar disk and since it is not possible to detect this case, the lifetime of some sunspot groups will be calculated as incomplete.} of the sunspot group and total number of days observed (column \quotes{DCnt}), respectively. If a sunspot group starts to appear on the solar disk before the first observation day of the year in question, a \quotes{started before} remark is put in its Comment column. Also, if a sunspot group is continuing to survive after the last observation of the year in question, similarly, a \quotes{group alive} remark is put in its Comment column. There is one thing to emphasis for the column of the {\it Group Evolution} in the list, in order to show the group evolution properly, three symbols (-), (*) and (X) beside group info are used to show the situations \quotes{no observation}, \quotes{day of backside} and \quotes{group not observed}, respectively. A sample for this case is given on the bottom part of Fig. 10 and in the middle part of Table 1, respectively\footnote{These samples were taken from the observatory annual report for 2012 which is being prepared}. 

\renewcommand{\arraystretch}{1.0} 
\begin{table*}[!b]  %%%%%%%%%%%%%%%%%% TABLE 1
	\caption{Some samples for the evolution of sunspot groups observed in 2012, which obtained by using SCAT. Symbols (-), (*)  and (X) in the group evolution column represent the days \quotes{no observation}, \quotes{day of backside} and \quotes{group not observed}, respectively.}
	\label{T-evolution-2012}
	\small
	\begin{tabular}{@{}crrrrp{11.2cm}@{}}
		\tableline
		&\multicolumn{2}{c}{Heliographic} & \multicolumn{2}{c}{Observation} & \\ 
		GNo & Lat. & Long. & First & Last & Group Evolution\\
		\tableline
		1 & -25 &  98 & I,02 &   I,05 & HS-1, CAI-7, CAO-4, HS-2 \\
		2 &  12 & 200 & I,16 &   I,24 & HA-1, HA-1, HA-2, HS-2, -, -, AX-4, CAI-5, AX-1 \\
		3 & -19 & 147 & I,02 &   I,04 & HS-1, HA-3, HA-1 \\
		4 &  10 & 121 & I,02 &   I,05 & CSI-16, CRI-16, BXI-35, CAO-7 \\
		5 & -18 &  86 & I,02 &   I,05 & HS-4, HK-12, HK-8, HS-2 \\
		\vdots & \vdots&\vdots& \vdots & \vdots & \hspace*{1.5cm} \vdots \hspace*{2.5cm} \vdots \\ [1ex]
		238 &  13 & 218 &  VI,27 &  VI,29 & AX-1, X, AX-2 \\
		239 & -15 & 218 &  VI,27 & VII,03 & BXI-8, BXI-13, BXI-16, BXI-24, BXI-5, BXI-6, BXI-5 \\
		240 & -17 & 209 &  VI,27 & VIII,27 & HA-2, DSO-4, DAC-15, EKC-24, EKC-41, EKC-55, FSC-54, EKC-37, FKC-62, FKC-43, FKC-25, FAC-16, ?-3, *, *, *, *, *, *, *, *, *, *, *, *, *, *, ?-3, CAO-3, CAO-3, CSO-2, CSI-5, HA-5, CSO-2, HH-2, HS-1, HS-1, HS-1, HS-1, X, *, *, *, *, *, *, *, *, *, *, *, *, *, X, ?-1, HS-1, HS-4, HS-1, HA-1, HA-1, HR-1, AX-2 \\
		241 &  17 & 214 &  VI,28 &  VI,29 & HR-3, HA-4 \\
		242 &  14 & 205 &  VI,28 & VII,02 & AX-2, BXO-10, BXO-12, BXI-10, BXI-5 \\
		243 &   8 & 307 &  VI,29 &  VI,29 & AX-1 \\
		\vdots & \vdots &\vdots& \vdots &  \vdots & \hspace*{1.5cm} \vdots \hspace*{2.5cm} \vdots \\ [1ex]
		461 & -16 & 308 & XII,27 & XII,31 & ?-6, CAO-5, -, BXI-6, AX-3 \\
		462 &   6 &  60 & XII,28 & XII,28 & AX-2 \\
		463 & -16 & 323 & XII,28 & XII,28 & AX-2 \\
		464 &  27 & 319 & XII,30 & XII,31 & BXO-4, CAI-9 \\
		465 &  14 & 338 & XII,31 & XII,31 & BXO-6 \\
		466 &   3 & 273 & XII,31 & XII,31 & BXO-4 \\
		467 & -24 & 250 & XII,31 & XII,31 & ?-1 \\		\tableline
	\end{tabular}
\end{table*}

In Sec-B, the number of sunspot groups and umbrae along with the observer’s initials are listed for each observation day of the year. The same listing is performed for Sec-C to show the relative number for each day. In Sec-D, the folder’s cover where all observation papers are held in is prepared. In Sec-E, the all data to prepare the table and figures in the observatory annual report are shown as both text output and picture. Finally in Sec-F, all observation days of the year in question can be checked for both position and group number of the sunspot groups visually. These descriptions will be better understood as a whole with sample outputs shown in the next section.

\section{The Outputs of the Year 2012 as a Sample}
At the Istanbul University Observatory, the number of observation days in the year of 2012 is 315 and total number of the sunspot groups observed is 2054, which is the sum of daily observed sunspot group numbers. After the group numbering process of the year 2012 is completed, the obtained number of the sunspot groups that has a lifetime one day or longer is 467, and some selected results of the sunspot group evolution are listed in Table 1 briefly. Here, an example of a sunspot group that makes three revolutions around solar disk is specially given to show the capability of the method. During the process in Sec-A, while the group evolution is extracted, the latitudinal distribution of the sunspot groups and the count and percentage of group types are also obtained simultaneously. Then, the results of this process are shown in Sec-E both graphically and as text output. Group types distribution is shown in Fig. 11 and the output window of Sec-E is shown in Fig. 12, respectively.

\begin{figure}[!t!]\label{F-figures-11}
	\centerline{\includegraphics[width=1.0\columnwidth]{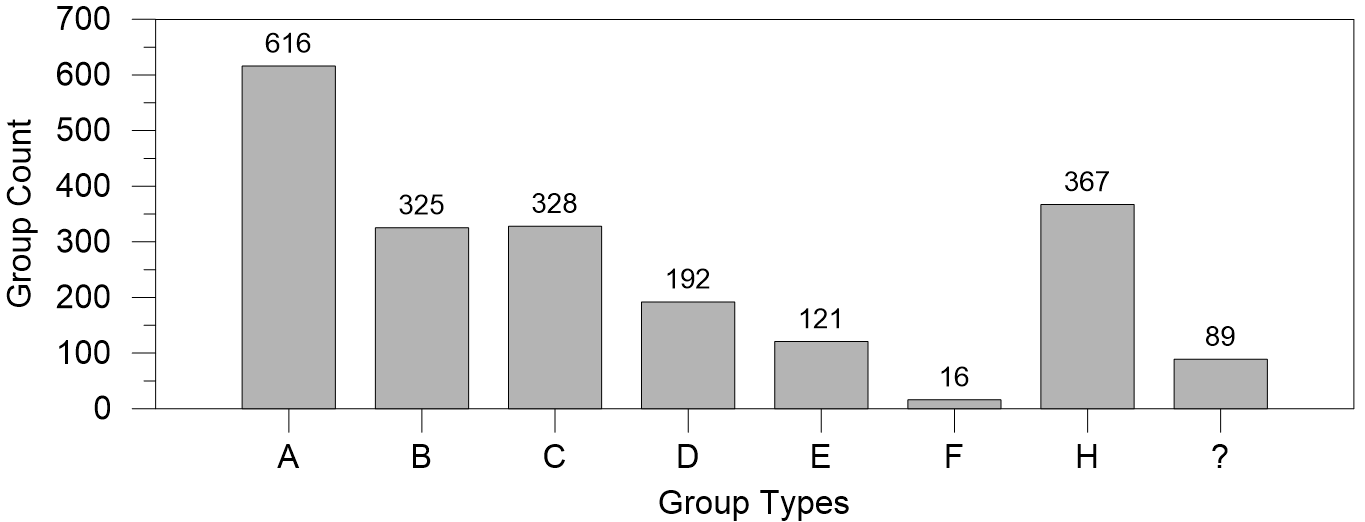}}
	\caption{Group types distribution in the year 2012 as a sample graphic obtained from Sec-E.}
\end{figure}

The outputs from the Sec-B and Sec-C that shown side by side in Fig. 13 are also taken as text outputs to prepare tables and figures for observatory annual report. Additionally, it is possible to get the development of sunspot relative number both monthly and yearly with these outputs. As seen from the right panel of Fig. 12, it is also possible to extract the needed data from these outputs for different purposes. For example, the latitudinal distribution of the sunspot groups can be obtained for different latitude ranges, which can be 1 degree width as well as 5 degree width.

\begin{figure*}[!b!]\label{F-figures-12}
	\vspace*{5mm}
	\centerline{\includegraphics[width=0.85\textwidth]{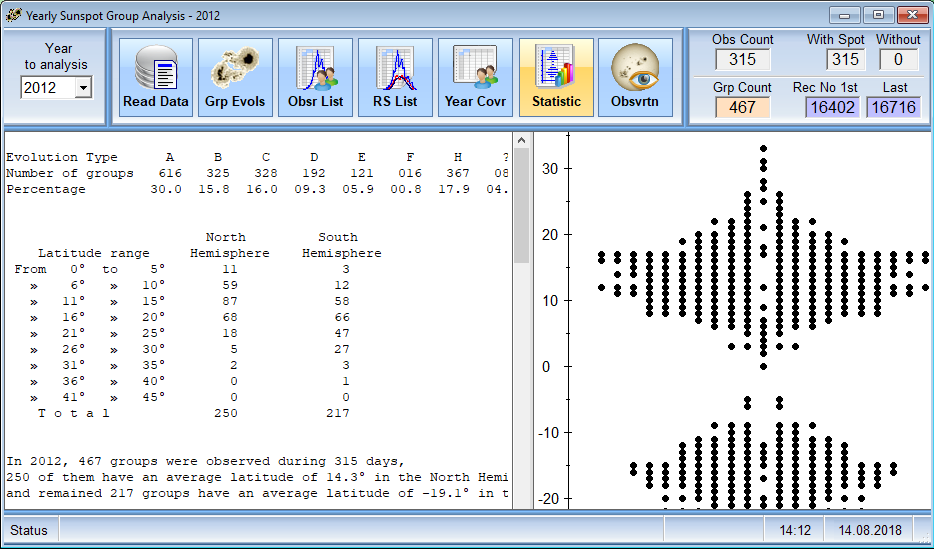}}
	\caption{Two output samples of Sec-E that are obtained during the process of Sec-A. {\it Left}) Text outputs to prepare many graphics for the annual report, and {\it Right}) a graphical preview of the latitudinal distribution.}
\end{figure*}

\section{Discussion} %-------------------------------------------------
\label{S-discussion}  
Solar cycles show basically different developments from each other and none of them are similar. When each solar cycle is analyzed in this respect, one can find that every cycle has distinctive development. Therefore, revealing the differences in each cycle becomes an important milestone on the studies of solar cycle properties. Also, this makes it possible to obtain some clues about interior dynamics \citep{Parker1970, DikChar1999, KhoColl2008} of inner solar layers during magnetic activity. Maybe, some signs for cyclical changes in solar differential rotation \citep{Beck1999} can be found by analyzing the evolution of long-lasting sunspot groups depending on a cycle's phase. Of course, SCAT can not find any solution to these issues directly. But it can contribute by combining the results of every cycle year to complement each other. For example, the sunspot relative number is calculated for each day of the month as shown in the outputs of Sec-C, then the monthly average sunspot relative number is calculated from these values easily. After finishing the analysis of each year of the cycle, the general changes of the sunspot relative number can be obtained for the whole cycle. Also, some special properties of a cycle can be revealed by comparing the group areas and lifetimes of the sunspot group evolution with each other observed on the rising and falling branches of the cycle. These comparison points can be increased with the obtained various data depending on the interested area.

\begin{figure*}[!t!]\label{F-figures-13}
	\vspace*{5mm}
	\centerline{\includegraphics[width=0.85\textwidth]{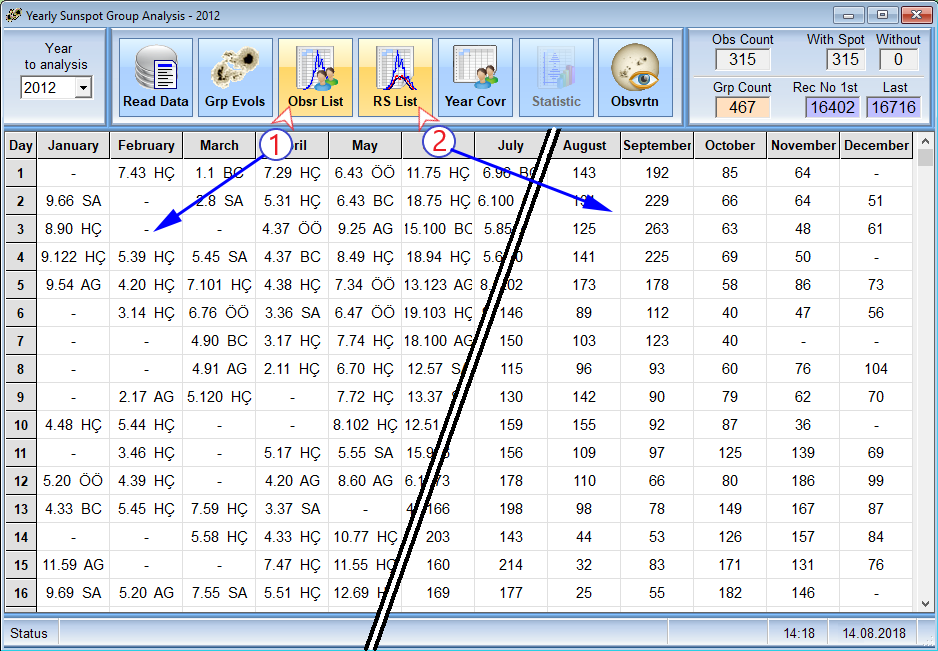}}
	\caption{Outputs of Sec-B and Sec-C for the year 2012. \textbf{1)} Sunspot group and umbra number along with the observer's initials, and \textbf{2)} sunspot relative number for each day of the year.}
\end{figure*}

The most important point in revealing sunspot group evolutions is to separate them properly. But this stage is mostly not as easy as thought. Naturally, the separation process varies from person to person because of the selected or accepted criteria for the group separation. One can accept the distribution of magnetic polarity as others can accept the space between the sunspot groups. The weather conditions also play an important role in observations and therefore in separating groups. Since all observers in an observatory are trained in that observatory's style of grouping sunspots, they will adopt that style. So it can be said that grouping in an observatory will be consistent over time and between observers. From this perspective, everyone can find a consistent solution for themselves. But which one will be correct, it is a subject of discussion and it will remain as a central problem in this subject.

As mentioned in the previous sections, the current state of SCAT is designed according to today's needs of our observatory and most of the obtained data are used to generate the tables and figures of the observatory annual report. On the other hand, as the developed method has a modular structure that can be easily modified, the future developments can be performed according to the requirements of observatory's studies on this subject. Importantly, in the near future, a specific study is planned to make this method open for everybody's use via a web-based application to give an opportunity to everyone to compare their data to other researchers' data.

\section{Acknowledgments}
Thanks to Assoc. Prof. Dr. Nurol Al from the Department of Astronomy and Space Sciences, Faculty of Science, Istanbul University for the idea to prepare this article.  Also, thanks to Res. Asst. Ba{\c s}ar Co{\c s}kuno{\u g}lu for his contributions towards improving the language of the manuscript. Also, thanks to the anonymous reviewer for his/her valuable suggestions and comments improving the manuscript. This research made help of Istanbul University Observatory Sunspot Observations. This work supported by the Istanbul University Scientific Research Projects Commission with the project number 24242.

%\nocite{*}
%\bibliographystyle{spr-mp-nameyear-cnd}
%\bibliography{myref}
%\bibliography{biblio-u1}

\begin{thebibliography}{}

\bibitem[Balthasar et al.(1986)]{Balthasar1986}
Balthasar, H., Vazquez, M. and Wöhl, H., 1986, \aap, 155, 87-98

\bibitem[Beck(1999)]{Beck1999}
Beck, J.G., 1999, \solphys, 191, 47–70

\bibitem[Bhatnagar \& Livingston(2005)]{BhatLiv2005}
Bhatnagar, A. and Livingston, W., 2005, Fundamentals of Solar Astronomy, World Scientific Series in Astronomy and Astrophysics - Vol. 6, p.213-216, World Scientific Publishing Co. Pte. Ltd

\bibitem[Clette(2011)]{Clette2011}
Clette, F., 2011, \jastp, 73, 182

\bibitem[Cristo \& Sánchez-Bajo(2011)]{Cristo2011}
Cristo, A., Vaquero, J.M. and Sánchez-Bajo, F., 2011, \jastp, 73, 187

\bibitem[{\c C}akmak(2014)]{Cakmak2014}
{\c C}akmak, H., 2014, \exas, 38, 77-89

\bibitem[Dikpati \& Charbonneau(1999)]{DikChar1999}
Dikpati, M. and Charbonneau, P., 1999, \apj, 518, 508-520

\bibitem[Green(1986)]{Green1986}
Green, R.M., 1986, Textbook on Spherical Astronomy, Reprinted, p.17, Cambridge University Press, Cambridge 

\bibitem[Howard et al.(1984)]{Howard1984}
Howard, R., Gilman, P.I. and Gilman, P.A., 1984, \apj, 283, 373-384

\bibitem[Khomenko \& Collados(2008)]{KhoColl2008}
Khomenko, E. and Collados, M., 2008, \apj, 689, 1379-1387

\bibitem[Lang(2009)]{Lang2009}
Lang, K.R., 2009, The Sun from Space Second Edition, p.75, Springer-Verlag Berlin Heidelberg

\bibitem[McIntosh(1990)]{McIntosh1990}
McIntosh, P., 1990, \solphys, 125, 251-267

\bibitem[Meadows(2002)]{Meadows2002}
Meadows, P., 2002, J. Br. Astron. Assoc., 112, 353-356

\bibitem[Parker(1970)]{Parker1970}
Parker, E.N., 1970, \apj, 160, 383-404

\bibitem[Waldmeier(1947)]{Waldmeier1947}
Waldmeier, M., 1947, Publ. Zürich Obs., 9, 1

\end{thebibliography}
%
%\end{document}

\end{document}